\documentclass[twocolumn,showpacs,preprintnumbers,amsmath,amssymb,prb,aps]{revtex4-2}

\usepackage{subfigure}
\usepackage{comment}
\usepackage{esvect}
\usepackage[usenames, dvipsnames]{color}
\usepackage{epsfig}
\usepackage{graphicx}
\usepackage{float} 
\usepackage{gensymb}
\usepackage{multirow}
\usepackage{dcolumn}   
\usepackage{bm}        
\usepackage{ulem}
\begin{document}
	
\title{Evidence of charge transfer in a bilayer film, La$_{0.67}$Sr$_{0.33}$MnO$_3$$/$YBa$_2$Cu$_3$O$_7$}
	
\author{Ankita Singh, Sawani Datta, Ram Prakash Pandeya, Srinivas C. kandukuri, and Kalobaran Maiti}
\altaffiliation{Corresponding author: kbmaiti@tifr.res.in}
\affiliation{Department of Condensed Matter Physics and Materials Science, Tata Institute of Fundamental Research, Homi Bhabha Road, Colaba, Mumbai-400005, India.}
	
	
\begin{abstract}
We study the growth and electronic properties of a high temperature superconductor, YBa$_2$Cu$_3$O$_7$ (YBCO) in proximity of a magnetic material, La$_{0.67}$Sr$_{0.33}$MnO$_3$ (LSMO). High quality single crystalline films of YBCO and LSMO/YBCO were grown epitaxially on SrTiO$_3$ (001) surface. Magnetization data of the LSMO/YBCO bilayer exhibit ferromagnetic ordering with Curie temperature smaller than that of pure LSMO. Measurements at different field directions reveal emergence of an anisotropy at low temperatures with in-plane easy axis; the observed anisotropy is stronger in the superconducting region. Magnetization data of YBCO exhibit onset of diamagnetism at 86 K for the out-of-plane magnetic measurements while the in-plane measurements show onset at a slightly higher temperature of 89 K with much smaller moment. Interestingly, the onset of diamagnetism in LSMO/YBCO film remains at 86 K despite the presence of ferromagnetic LSMO layer underneath. The analysis of Ba 4$d$ and Y 3$d$ core level spectra suggest that the surface and bulk electronic structure in these systems are different; the difference is reduced significantly in the LSMO/YBCO sample suggesting an enhancement of electron density near the surface arising from the charge transfer across the interface, which is consistent with the magnetization data.
\end{abstract}
	
\maketitle
	
\section{Introduction}

Superconductivity and magnetism are two mutually exclusive phenomena as the zero resistance condition can completely block the penetration of magnetic lines of force through the materials as demonstrated in Meissner effect. Application of magnetic field destroys superconductivity beyond a critical field and the penetration of lines of forces called vortices are studied in the intermediate field regime extensively. While some materials show coexisting antiferromagnetic order and superconductivity \cite{pnictides}, it is not clear if both the behavior occupy the same phase space. In general, it is difficult to achieve both kinds of behaviour coexisting in the same bulk system and attempts are made to prepare materials in the form of heterostructures of thin films having competing interactions \cite{Bergeret, Hoppler, Sefrioui, Djupmyr}. The antagonistic behaviour of the materials may lead to exotic scenarios as the superconductivity tends to expel the magnetic lines of force limiting them within the magnetic film and/or the spin-polarized quasiparticle injection occurs across the interface \cite{soltan2004ferromagnetic}. In a typical superconductor, electrons of opposite momenta and spins form Cooper pairs, wherein ferromagnetic compound, exchange interaction tends to align magnetic spins in the direction of the applied magnetic field. Thus, the ferromagnetic layer influences the critical current and vortex pinning strength in the superconducting layer, while the Cooper pairs in the superconducting layer may renormalize the magnetic exchange interactions within the ferromagnetic layer.

Bilayer films of magnetic and superconducting materials are studied extensively in the literature \cite{Zhang, chaudhuri2023interface, wisser2021growth, chen2009flux}, and La$_{0.67}$Sr$_{0.33}$MnO$_3$ (LSMO) and YBa$_2$Cu$_3$O$_7$ (YBCO) are found to be good candidate materials for epitaxial growth with sharp interface due to their similar \textit{a/b} lattice parameters. Various studies show exotic phenomena such as domain wall superconductivity \cite{houzet2006theory}, triplet superconductivity \cite{Samal}, change in vortex behaviour \cite{keizer2006spin, bhatt2022correlation}, etc. In a study by Hong-Ye \textit{et. al.}\cite{hong2009vortex} on superconducting/ferromagnetic multilayers, an enhancement of critical current density is reported in the low field regime exhibiting the presence of strong vortex pinning effect due to the presence of spin-vortex interaction. In other studies  \cite{PrzyslupskiPRB, Przyslupski} of LSMO/YBCO superlattices, the presence of anisotropy in LSMO layers and exchange biasing effect are also observed \cite{PrzyslupskiPRB}. The charge career density significantly influences the properties of LSMO \cite{LSMO-Bindu} and YBCO \cite{YBCO}. Thus, the injection of spins from the ferromagnetic layer to the superconducting one in the heterostructure will break the time reversal symmetry of the cooper pairs due to excess magnetic moments and quasiparticle redistribution \cite{Jeon} leading to a suppression of superconducting transition temperature, {\textit T$_c$}. Evidently the properties of heterostructures involving magnetic and superconducting layers is an interesting area with emerging exoticity arising from their interaction across the interface which is an unknown paradigm. In the present work, we study the structural and magnetic properties of LSMO/YBCO bilayer film and compare the data with the pristine YBCO case. Our results reveal interesting evolution of the properties of YBCO layer due to the presence of ferromagnetic layer underneath and evidence of charge transfer across the interface.

\section{Experimental details}

YBCO and LSMO films were grown on SrTiO$_3$ (001) surface using a home-built ultra high vacuum (UHV) Pulsed Laser Deposition (PLD) system equipped with KrF excimer laser ($\lambda$ = 248 nm) set at 5 Hz frequency. We used commercially available targets of (La$_{0.67}$Sr$_{0.33}$MnO$_3$, YBa$_2$Cu$_3$O$_7$ and the target to substrate distance was maintained at 5 cms. LSMO layers were grown with laser fluence of 1.5 Jcm$^{-2}$, substrate temperature of 750 $\degree$C and oxygen partial pressure of 300 mTorr. The growth parameters for YBCO was set at 1.8 Jcm$^-2$, 850 $\degree$C and 400 mTorr, respectively. The thickness of the films was monitored by the number of laser pulses used for deposition; 3000 shots for LSMO layer ($\sim$20 nm) and 21000 shots for YBCO layer ($\sim$50 nm) in all the cases. The films were \textit{in-situ} annealed for 1.5 hours under 500 Torr oxygen pressure in the deposition chamber and then slowly cooled down to room temperature.

The crystallinity and epitaxial quality of the films were studied by x-ray diffraction (XRD) measurements. For the structural property of the films, we used $\theta$/2$\theta$ scan mode of Rigaku diffractometer. Pole figure data was recorded for the tilt angle, $\phi$ (0$\degree$ - 90$\degree$) and azimuthal angle, $\psi$ (0$\degree$ to 360$\degree$). Reciprocal space mapping is carried out for YBCO(119) Bragg peak. The magnetic measurements were performed using SQUID magnetometer (MPMS XL Evercool, Quantum Design). The analysis of the electronic structure of the thin film was carried out using x-ray photoelectron spectroscopy (XPS) equipped with R4000 Scienta detector and monochromatic Al $K\alpha$ source at an energy resolution of 0.4 eV.

\section{Results and Discussion}

\begin{figure}
\centering
\includegraphics[width=0.5\textwidth]{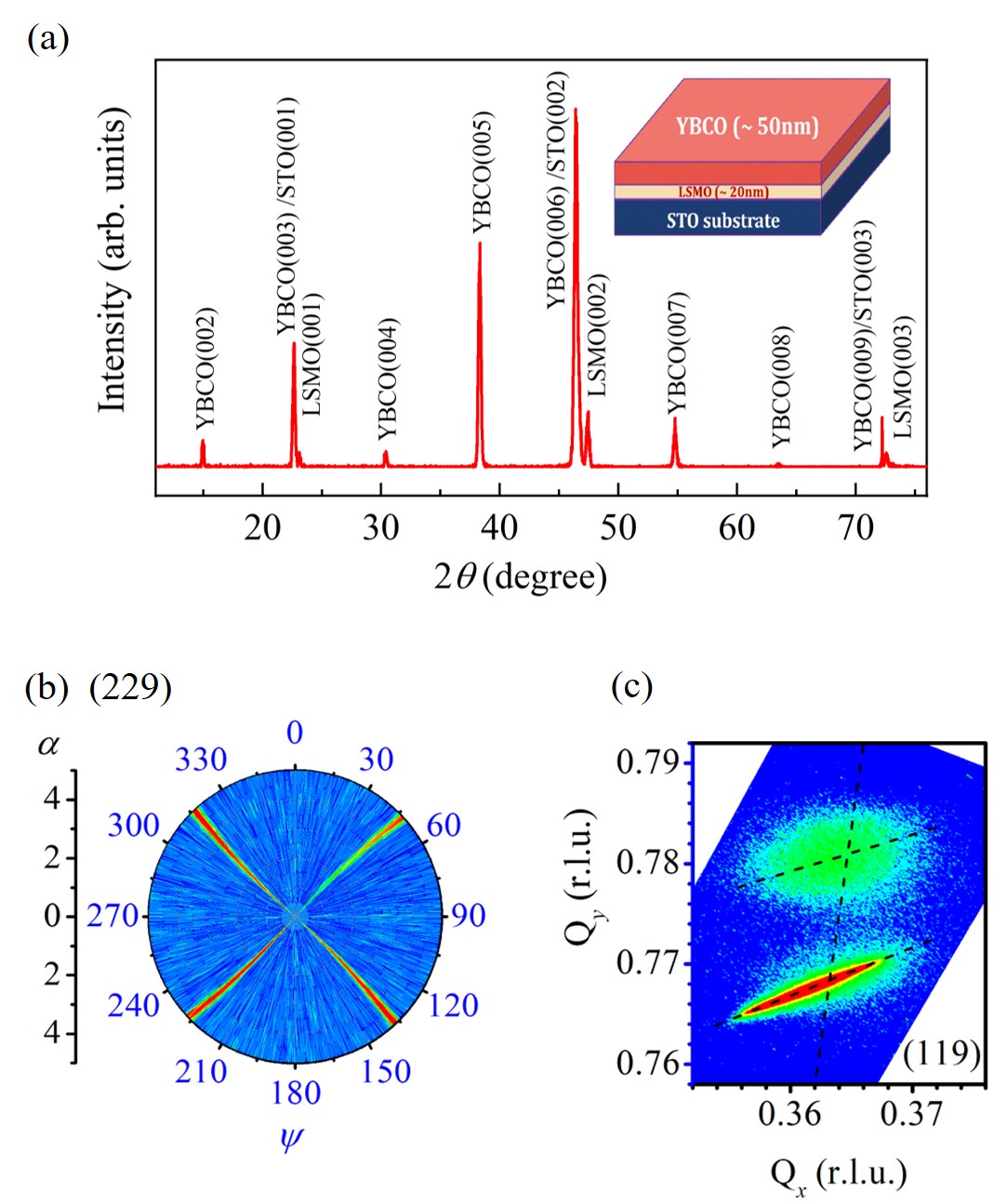}
\vspace{-2ex}
\caption{(a) XRD scan of LSMO/YBCO heterostructure; inset shows the schematic of the bilayer film studied.(b) Pole figure measurements along the in-plane axis of YBCO (229) for the bilayer LSMO/YBCO thin film. (c) Reciprocal Space Mapping along (119) asymmetric scan of YBCO layer showing STO and YBCO.}
\label{XRD}
\end{figure}

YBCO forms in orthorhombic structure having lattice parameters, \textit{a} = 3.82 \AA, \textit{b} = 3.89 \AA, and \textit{c} = 11.68 \AA\cite{lorenz2001high}. These \textit{a}, \textit{b} values are comparable to the lattice parameter of STO substrate (\textit{a} = 3.905 \AA) and pseudocubic lattice parameter of LSMO (\textit{a} = 3.87 \AA)\cite{chaudhuri2023interface}, which is good for the high quality epitaxial growth of both these layers on the STO substrate. We analysed the crystal structure of the films grown employing x-ray diffraction techniques. A typical XRD scan of the bilayer film is shown in Fig. \ref{XRD}(a). The schematic of the film is shown in the inset of the figure. No secondary phase is observed for YBCO and LSMO. The data indicate (00\textit{l}) oriented growth in all the cases suggesting good single crystallinity of the samples \cite{singh2023structural}.

In order to investigate the epitaxial quality and orientation of the films further, we performed pole figure measurements for the YBCO (229) reflection of the bilayer film. Since, the (229) reflection of YBCO was not found in the $\theta/2\theta$-scan, we explored the in-plane reflections for these measurements; the YBCO(229) plane appears at $\alpha$ (90$\degree$ - $\phi$) = 5$\degree$. The experimental pole figure data is shown in Fig. \ref{XRD}(b) with azimuthal angle, $\psi$ as the circumference axis and $\alpha$ as the radial direction at an expanded scale of 0-5$\degree$. The data exhibit four distinct features with high intensity at $\alpha$ = 5$\degree$ which is consistent the growth having (00$\textit{l}$)-direction as the out-of-plane axis.

The Reciprocal Space Mapping (RSM) is performed for (119) reflection of YBCO as shown in Fig. \ref{XRD}(c). It is evident that the broadening axis of YBCO makes an angle with the $Q_x$ direction and has elliptical shape. This confirms the presence of mosaic structure i.e. the thin film consists of single crystal blocks whose in-plane axis are not in line with each other \cite{chen2000influence, liu2008microstructure}. We calculated the lateral correlation length and microscopic tilts that provides information about the average block size and out-of-plane misorientation, respectively which results into the elliptical shape. The calculated lateral correlation length and microscopic tilt for YBCO is found to be 187.72 nm and 0.018$\degree$, respectively, which confirms high quality epitaxial growth along (00\emph{l}) direction. The center of the substrate and the YBCO layer does not lie in the same vertical line suggesting a relaxed growth of the YBCO film. The in-plane and out-of-plane lattice parameters are estimated to be 3.92 \AA and 11.84 \AA, respectively which are close to the bulk lattice parameters.

\begin{figure}
\centering
\includegraphics[width=0.5\textwidth]{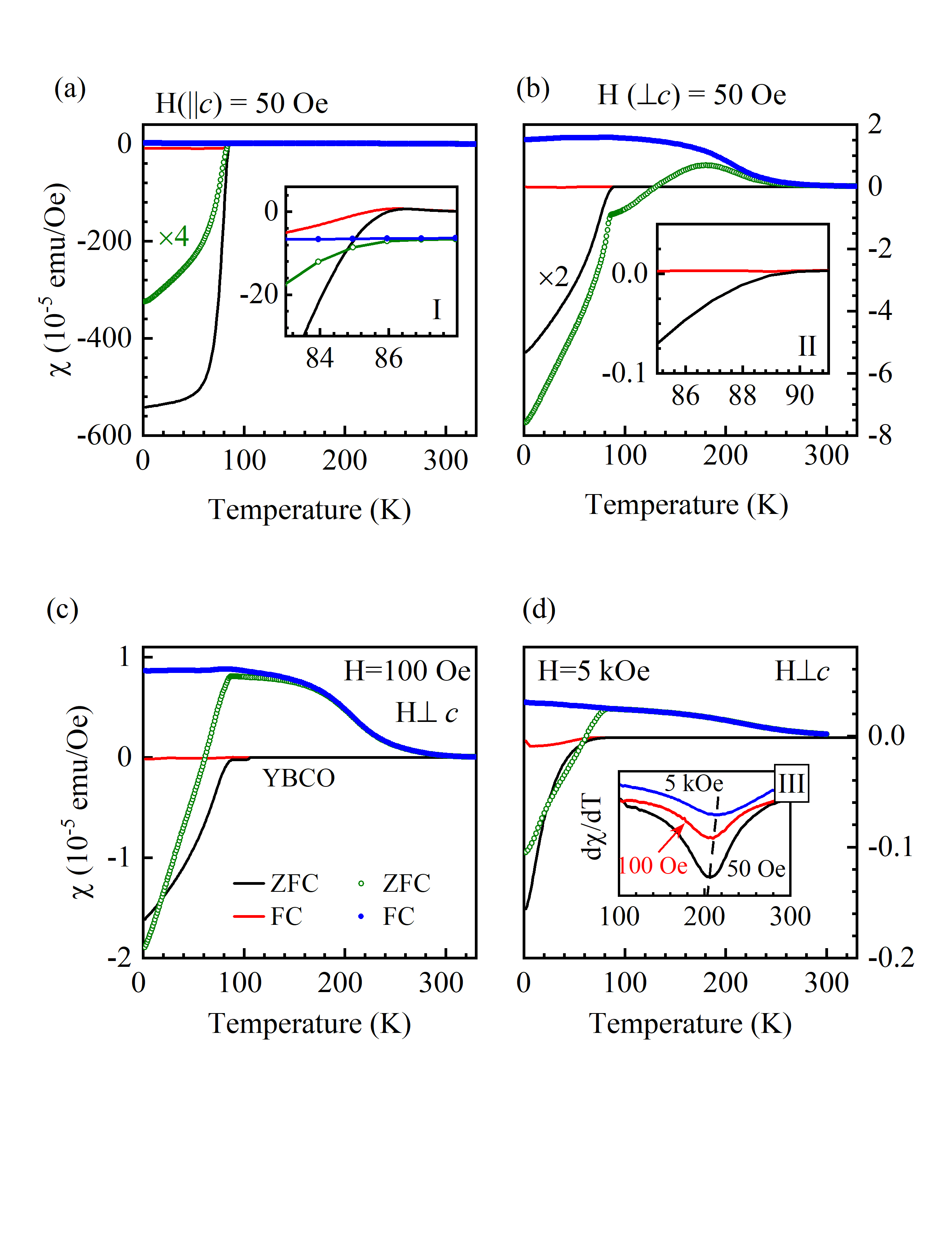}
\vspace{-18ex}
\caption{FC and ZFC magnetization data for 50 Oe external field (a) perpendicular to $c$-axis and (b) parallel to $c$-axis. The symbols represent data for LSMO/YBCO film and the lines represent the data for YBCO film. Inset I shows the data in (a) in an expanded scale and the data for LSMO/YBCO sample is shifted down by 8 units for clarity. Inset II shows the YBCO data in (b) in an expanded scale. The magnetization data perpendicular to $c$-axis for the field of (c) 100 Oe and (d) 5 kOe. The inset III shows the derivative of magnetization vs temperature data for 50 Oe, 100 Oe, and 5 kOe applied field perpendicular to the $c$-axis exhibiting variation of the Curie temperature of LSMO/YBCO sample with the applied field.}
\label{Fig2-MT}
\end{figure}

The magnetization measurements (zero field cooled, ZFC and field cooled, FC) were performed in a temperature range of 2-350 K under an applied magnetic field of 50 Oe, 100 Oe and 5 kOe. In Fig. \ref{Fig2-MT}, the lines show the experimental data for YBCO film and the symbols represent the magnetization data of LSMO/YBCO film. In Fig. \ref{Fig2-MT}(a), we show the data collected at 50 Oe magnetic field along the $c$-axis; the data for LMSO/YBCO sample is multiplied by 4. Onset of superconducting phase is observed at about 86 K; this is shown in Inset I with expanded $x$- and $y$-scales. The LSMO/YBCO data are shifted by 2 emu/Oe towards negative $y$-axis for clarity in the figure. While the diamagnetic transition in the ZFC-curve is sharp and intense for YBCO, it is less pronounced in the LSMO/YBCO sample. Interestingly, the onset of superconductivity occurs at the same temperature in LSMO/YBCO case despite the presence of ferromagnetic layer underneath.

In Fig. \ref{Fig2-MT}(b), we show the magnetization data for the in-plane field of 50 Oe ($\bot c$-direction). The ZFC curve of the YBCO film show onset of superconductivity at a slightly higher temperature ($\sim$ 89 K) with the diamagnetic moments significantly less than the values observed for out-of-plane magnetization direction. The data for the bilayer film exhibit a ferromagnetic transition at about 255 K which is significantly smaller than the bulk Curie temperature of 360 K\cite{LSMO-bulk} and the 340 - 350 K observed in films of thickness 29 - 40 nm \cite{Balint-PNAS, Tulapurkar}. The reasons for the suppression of Curie temperature may be related to the structural modification to 2D film on a substrate of slightly different lattice parameters, imperfections in the materials \cite{CaB6}, and/or transfer of charge carriers between YBCO and LSMO layers \cite{LSMO-Bindu,YBCO}. Interestingly, distinct transition to the superconducting phase is observed at 86 K in the ZFC data of LSMO/YBCO sample; the transition temperature is similar to that for field along out-of-plane direction. Moreover, the diamagnetic moment for in-plane magnetization is found to be stronger than the pristine case.

The magnitude of diamagnetic signal for out-of-plane magnetization is significantly larger than the in-plane case presumably due to the large screening effect. Another important difference between in-plane and out-of-plane data is observed in the FC curve exhibiting a reduction in magnetization in out-of-plane direction whereas the in-plane FC curve increases down to the lowest temperature studied. This suggests that diamagnetism has dominant effect for the out-of-plane magnetization and the magnetic easy axis is in-plane of LSMO/YBCO layer as expected due to shape anisotropy. Slightly smaller superconducting transition temperature, $T_c$ in LSMO/YBCO relative to YBCO may be attributed to the following scenarios: (i) magnetic field due to ferromagnetic LSMO layer destroys the Cooper pairs and/or superconducting coherence close to the transition although eventually superconductivity sets in at a slightly lower temperature. (ii) Charge transfer across the interface \cite{chaudhuri2023interface}. Interestingly, onset temperature of superconductivity in both the samples, LSMO/YBCO and YBCO films, is identical for the out-of-plane field.

In Fig. \ref{Fig2-MT}(c) and (d), we show the ZFC and FC data of YBCO and LSMO/YBCO films at the in-plane magnetic fields of 100 Oe and 5 kOe. In the YBCO film, higher applied field reduces the diamagnetic susceptibility, $\chi$ confirming gradual destruction of Meissner effect with increasing applied magnetic field. $T_c$ of YBCO film reduces significantly with the application of higher field; the bifurcation of FC/ZFC curves occurs at 85 K and 65 K for 100 Oe and 5 kOe fields, respectively. In the LSMO/YBCO film, $\chi$ is positive at higher temperatures. The FC/ZFC curves bifurcate at about 84 K and 76 K for the field 100 Oe and 5 kOe, respectively. While reduction of $T_c$ due to application of magnetic field is in line with the expected behavior, the $T_c$ at 5 kOe appears to be higher in LSMO/YBCO film compared to the pristine case. The inset of Fig. \ref{Fig2-MT}(d) shows the variation of first derivative of susceptibility with the temperature in the vicinity of ferromagnetic Curie temperature. The minima of the first derivative curve appears at 206 K, 210 K and 216 K for 50 Oe, 100 Oe and 5 kOe, respectively indicating enhancement of ferromagnetic ordering temperature with the increase in field. Application of stronger magnetic field usually helps to align the moments better without much effect on the ordering temperature. However, enhancement of the ordering temperature in this system suggests that YBCO layer has significant influence exchange coupling within the LSMO layer even when the material is not superconducting, thereby, changes the magnetic ordering temperature.

\begin{figure}
\centering
\includegraphics[width=0.5\textwidth]{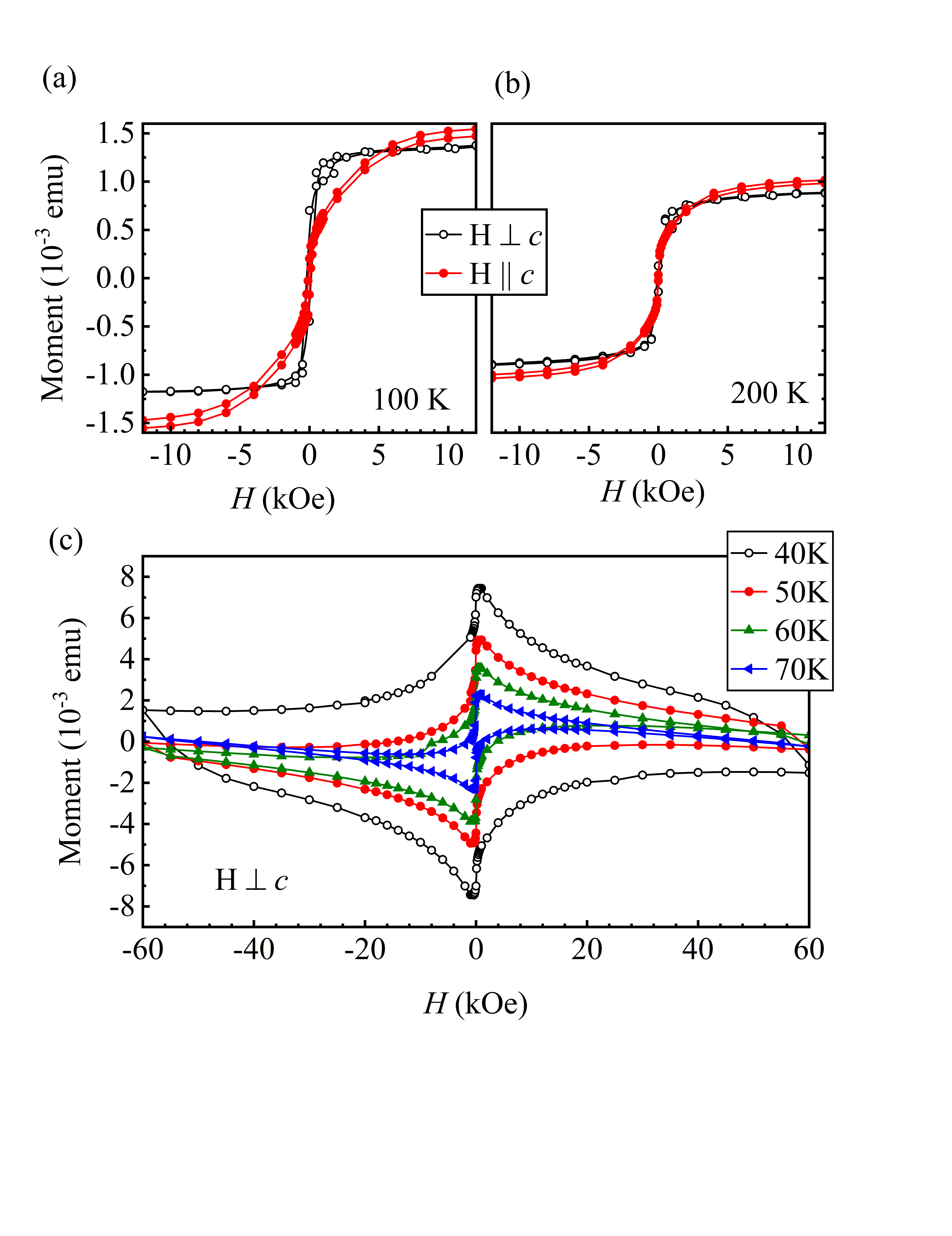}
\vspace{-18ex}
\caption{Magnetization loops of LSMO/YBCO bilayer film for the magnetic field applied parallel (closed circles) and perpendicular (open circles) to $c$-axis at (a) 100 K and (b) 200 K. (c) Magnetization loops at 40, 50, 60, and 70 K for the in-plane external field.
}
\label{Fig3-MH}
\end{figure}

Magnetic hysteresis loop of the LSMO/YBCO bilayer measured at 100 K and 200 K are shown in Fig. \ref{Fig3-MH}(a) and (b), respectively. The data at 200 K exhibit almost similar behavior with saturation moment for out-of-plane direction slightly larger than the in-plane moment. At 100 K, the results show significantly different behavior with emergence of large magnetic anisotropy at higher fields. While the area inside the loop is small in both the cases, the saturation moment is larger in the out-of-plane direction. The in-plane magnetization exhibit faster saturation than the out-of-plane magnetization.

In Fig. \ref{Fig3-MH}(c), we show the magnetic hysteresis loops at 40K - 70K for LSMO/YBCO bilayer for the in-plane fields. The hysteresis loops show complex behaviour exhibiting an interplay between Meissner currents in the YBCO layer and the ferromagnetic moments in the LSMO layer. While the behavior indicates presence of diamagnetism, the hysteresis loops exhibit plateau at higher field indicating signature of antiparallel alignment of the in-plane magnetic moments. This suggests strong proximity effect of the YBCO layer on the magnetism of LSMO layer presumably due to charge transfer across the interface as well as the presence of Cooper pairs\cite{kumawat2023magnetic}. From the phase diagram of La$_{1-x}$Sr$_x$MnO$_3$ \cite{Przyslupski}, it is known that the material show antiferromagnetic phase for doping level $x >$ 0.48. Thus, in the bilayer film, few LSMO layers near the interface possibly have gone into an antiferromagnetic state due to charge transfer \cite{Przyslupski, PrzyslupskiPRB}. While the behavior remain almost similar, the moments enhances with the decrease in temperature suggesting enhanced diamagnetism due to the superconductivity of the YBCO layers.

\begin{figure}
\centering
\includegraphics[width=0.5\textwidth]{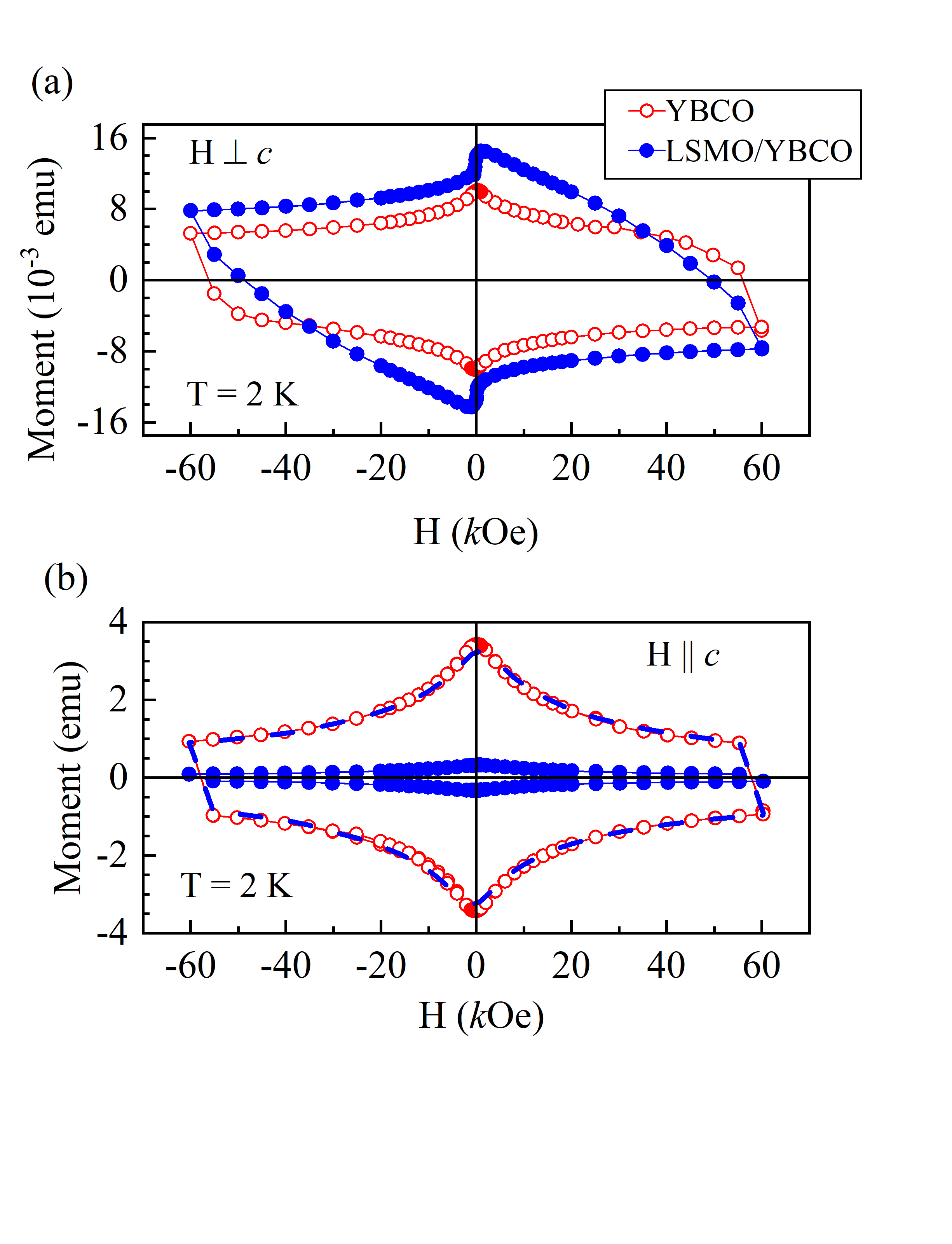}
\vspace{-18ex}
\caption{Magnetization loops of the YBCO film and YBCO/LSMO bilayer for the magnetic field applied (a) perpendicular (in-plane) and (b) parallel (out-of-plane) to $c$-axis direction at sample temperature of 2 K. Closed circles show the data for LSMO/YBCO bilayer and the open circles are the data for YBCO film. the dashed line in (b) represent the LSMO/YBCO data scaled by 10 times showing shape almost identical to that found for YBCO film.
}
\label{Fig4-MH_2K}
\end{figure}

Magnetic hysteresis loops measured at 2 K are shown in Fig. \ref{Fig4-MH_2K} for in-plane and out-of-plane magnetization directions. The data for YBCO and LSMO/YBCO are shown by open and closed circles, respectively. For in-plane magnetization, the data shown in Fig. \ref{Fig4-MH_2K}(a)exhibit a slightly different magnetic moment for YBCO and LSMO/YBCO cases. While YBCO film show a typical behavior for a superconducting material, the hysteresis loop for LSMO/YBCO film indicates a change in shape due to the presence of ferromagnetic moment. Here, the ferromagnetic behavior appears to be superimposed over the large superconducting layer's Meissner effect.

The hysteresis loops in the out-of-plane direction are shown in Fig. \ref{Fig4-MH_2K}(b). The behavior of the YBCO film is almost similar to the in-plane magnetization case with a slightly reduced moment. The data for both the samples exhibit no change in the shape of hysteresis loop though the magnetization for bilayer film is an order of magnitude smaller than pristine YBCO film. This is verified by plotting the rescaled LSMO/YBCO data by 10 times (dashed line) in the figure. Such strong reduction may be due to the presence of ferromagnetic LSMO layer. It appears that the magnetic moment of LSMO layer compensate the diamagnetism significantly resulting in the smaller overall magnetization value. These results suggest that the spins of the electrons in the superconducting phase may be aligned in the out-of-plane direction.

Previously, transfer of holes has been observed between the cuprates and manganites using optical far-infrared spectroscopy. Also, in La$_{0.67}$Sr$_{0.33}$MnO$_3$(LCMO)/YBCO superlattices, polarized neutron reflectometry studies show suppression of magnetization at the LCMO and YBCO interface. In addition to this, the transformation of few layers of ferromagnetic manganite into antiferromagnetic state due to the hole charge transfer is also shown using neutron reflectometry experiment \cite{Laverde}. Evidently, the behavior of the bilayer film observed here can have a strong influence from the charge transfer across the interface leading to a larger doped regime close to the interface.

\begin{figure}
\centering
\includegraphics[width=0.5\textwidth]{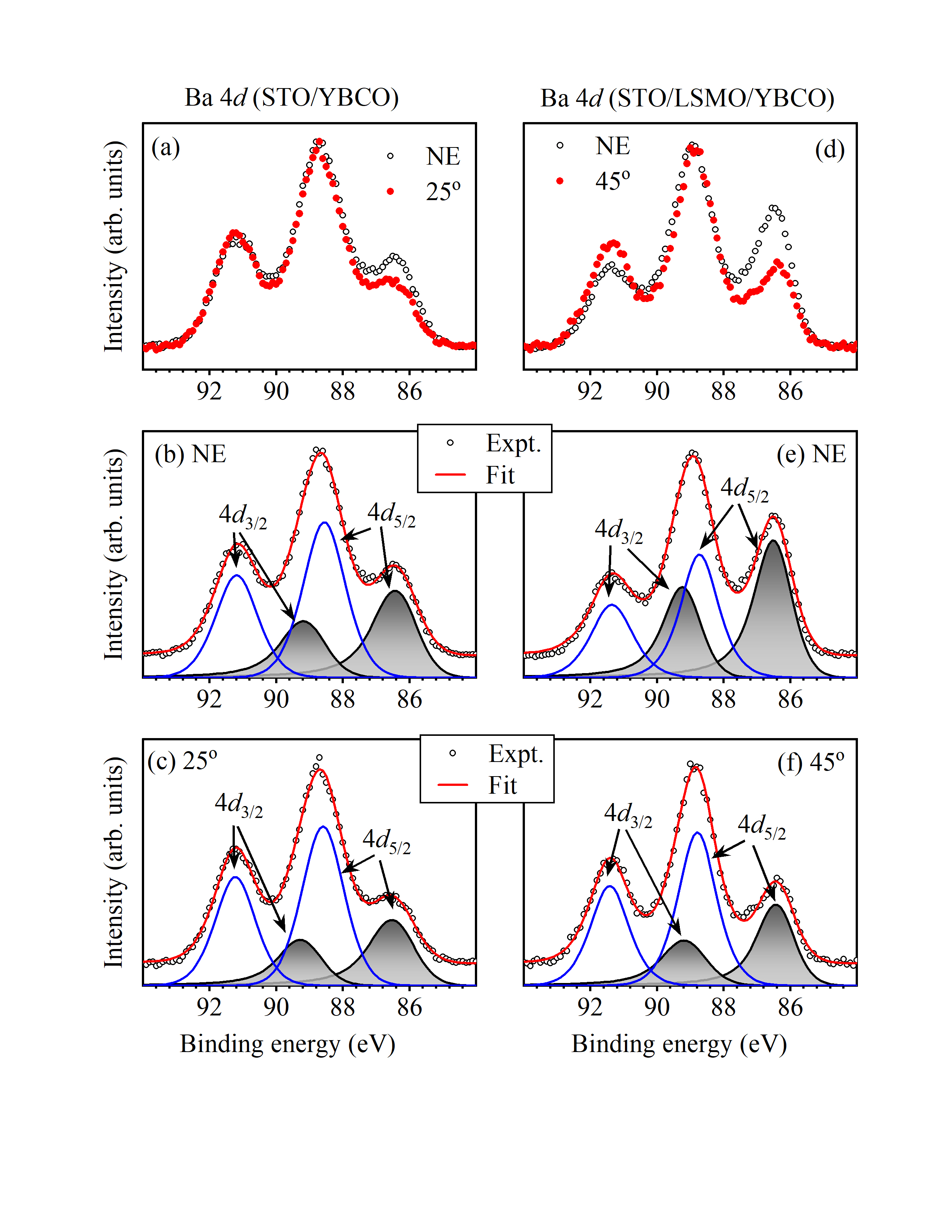}
\vspace{-8ex}
\caption{(a) Ba 4$d$ spectra of YBCO film at normal emission (open circles) and 25$^o$ angled emission (closed circles) geometry.  The simulation of these spectra for (b) normal emission and (c) 25$^o$-angled emission cases; the component peaks are shown at the bottom panel.
(d) Ba 4$d$ spectra of LSMO/YBCO film at normal emission (open circles) and 45$^o$ angled emission (closed circles) geometry.  The simulation of these spectra for (e) normal emission and (f) 45$^o$-angled emission cases; the component peaks are shown at the bottom panel.
}
\label{Fig5-Ba4d}
\end{figure}

In order to study the scenario further, we studied the Ba 4$d$ and Y 3$d$ core level spectra using XPS. These spectator elements have little contribution in the valence and conduction bands. However, the photoemission final state effect for the core level photoexcitation often manifests the changes in the electronic structure and hence provides a good way to study the signature of the charge transfer across the interface without involving the much complexity \cite{spectator}. Ba 4$d$ data collected from YBCO and LSMO/YBCO samples are shown in Fig. \ref{Fig5-Ba4d}. While Ba 4$d$ photoemission is expected to show 2 spin-orbit split features, we observe 3 distinct peaks. The spectra from YBCO sample are shown in Fig. \ref{Fig5-Ba4d}(a). A change in the photoemission geometry from normal emission (NE) to angled emission (25$\degree$-emission angle with respect to surface normal), the intensity of the peak at 86.4 eV binding energy reduces significantly with respect to the other features. The escape depth of the photoelectrons is sensitive to the emission angle; the 25$\degree$ angled emission will be more surface sensitive than the normal emission case \cite{surface}. The change in intensity with the change in surface sensitivity suggests that the peak at 86.4 eV binding energy is arising from bulk Ba sites while the features at higher binding energy are surface Ba-contributions. This description is consistent with the earlier studies of bulk single crystals \cite{YBCO-hike}. In order to find the signatures of the surface and bulk peaks in the spectral function, we have simulated the experimental spectra using a set of asymmetric Voigt functions representing the spin-orbit split features with intensity ratio based on the degeneracy of the features. The simulated data are shown in Figs. \ref{Fig5-Ba4d}(b) and (c) for NE and 25$\degree$ cases, respectively capturing the experimental spectral function excellently well. From this analysis, it is clear that the bulk and surface Ba 4$d_{5/2}$ signal appears at 86.4 eV and 88.6 eV binding energies and the spin-orbit splitting is found to be 2.8 eV. The binding energy of the bulk Ba-atoms is similar to the observation on bulk single crystals of YBa$_2$Cu$_3$O$_7$ \cite{YBCO-hike}. The Binding energy of the surface Ba-atoms is similar to the divalent Ba in BaO \cite{XPS-BaO}. While Ba is close to divalent in YBCO crystals, the surface Ba-atoms appears to be essentially divalent and the bulk Ba has some covalency as expected in solid systems.

The spectra from LSMO/YBCO are shown in Fig. \ref{Fig5-Ba4d}(d) for the NE and 45$\degree$ angled-emission cases exhibiting three distinct features similar to the YBCO case. Interestingly, the bulk feature in the bilayer data is significantly more intense than that in the YBCO case. This is curious considering the fact that the XPS data essentially represent the photoemission signal from top few layers of the YBCO film as the escape depth of the Be 4$d$ photoelectrons will be about 20 \AA\ \cite{surface}. The spectral change suggests significant change in the electronic structure due to the presence of LSMO layer below the YBCO layer.

\begin{figure}
\centering
\includegraphics[width=0.5\textwidth]{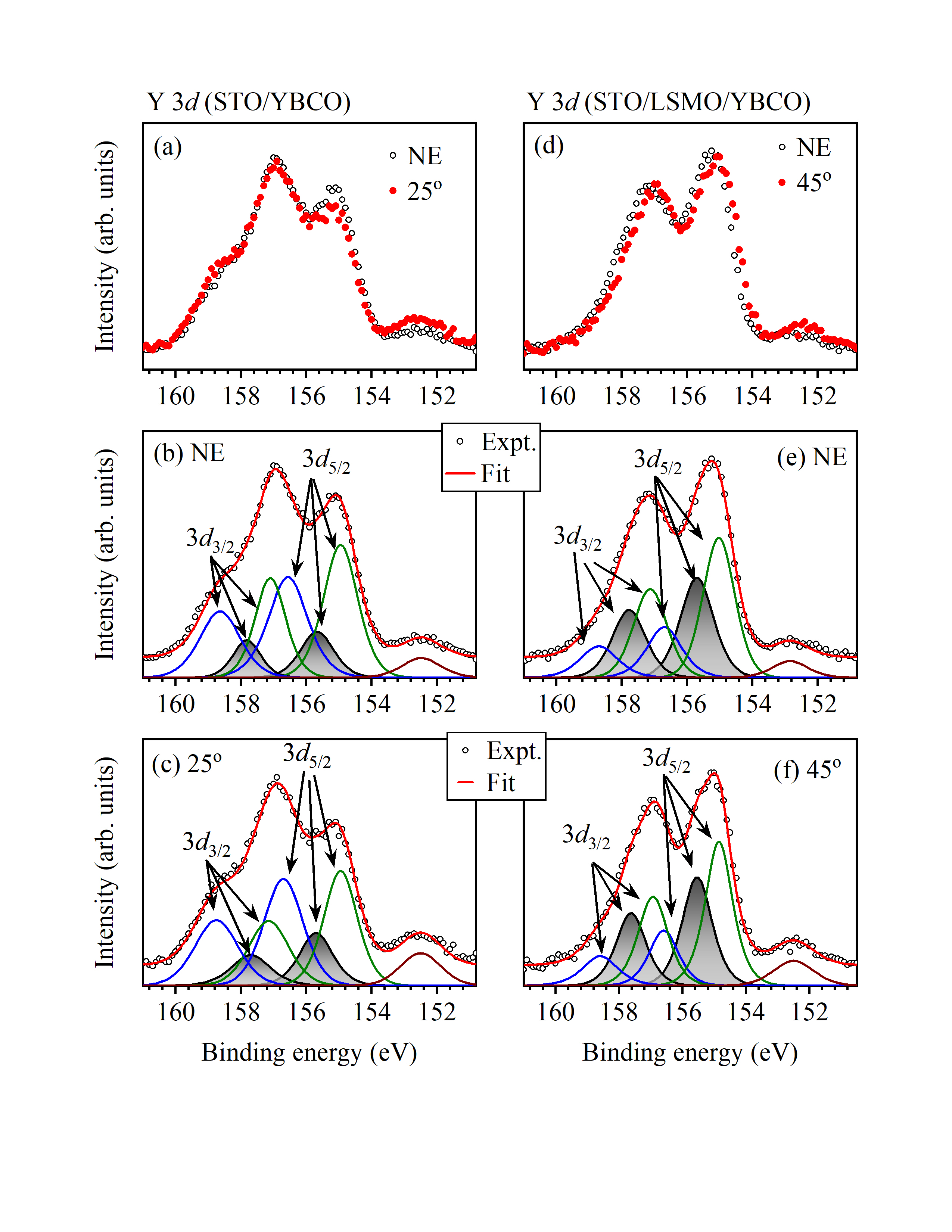}
\vspace{-8ex}
\caption{(a) Y 3$d$ spectra of YBCO film at normal emission (open circles) and 25$^o$ angled emission (closed circles) geometry.  The simulation of these spectra for (b) normal emission and (c) 25$^o$-angled emission cases; the component peaks are shown at the bottom panel.
(d) Y 3$d$ spectra of LSMO/YBCO film at normal emission (open circles) and 45$^o$ angled emission (closed circles) geometry.  The simulation of these spectra for (e) the normal emission and (f) 45$^o$-angled emission cases; the component peaks are shown at the bottom panel.}
\label{Fig6-Y3d}
\end{figure}		

In Fig. \ref{Fig6-Y3d}, we study the Y 3$d$ spectra collected at different emission angles. The spectra from YBCO sample shown in Fig. \ref{Fig6-Y3d}(a) exhibit 4 distinct features at 152.5, 155, 157 and 158.8 eV binding energies. This is unusual considering the photoemission studies in bulk YBCO crystals \cite{YBCO-hike, XPS-YBCO} where Y 3$d$ spectra often exhibit a two-peak structure for the spin-orbit split features. We notice that a change emission angle to 25$\degree$ that makes the technique more surface sensitive leads to a reduction in intensity around 155 eV. The feature at 152.5 eV may be an impurity feature and becomes stronger in the 25$\degree$-angled emission case suggesting it's link to the surface electronic structure. We have simulated the experimental spectra considering peaks representing spin-orbit split Y 3$d$ features; it was necessary to consider at least three sets of such features. The simulated spectra and the constituent peaks are shown in Figs. \ref{Fig6-Y3d}(c) and (d) for the NE and 25$\degree$-emission cases respectively. Y 3$d_{5/2}$ features in each set appears at 155 eV, 155.7 eV and 156.6 eV. The spin-orbit splitting is found to be about 2.2 eV which is similar to the earlier observations \cite{YBCO-hike, XPS-YBCO}. The peak at 156.6 eV may be attributed to trivalent surface yttrium while the peak at 155 eV is the bulk Y \cite{XPS-Y2O3}.

In Fig. \ref{Fig6-Y3d}(d), we show the Y 3$d$ spectra collected from LSMO/YBCO sample exhibiting a different scenario; a two peak structure as often observed in such systems. However, the relative intensity of the features does not match with their branching ratio. With the change in emission angle to
45$\degree$, the peaks appear to shift towards lower binding energy in contrast to the case for Ba photoemissions. The simulation of the experimental features exhibiting similar peak positions for the constituent peaks as found in YBCO sample while their relative intensities are very different. The intensity of the bulk Y peak becomes dominant in the LSMO/YBCO sample as also observed in the case of Ba photoemission. The surface Y-peak is significantly less intense in this case.

The overall reduction of intensity of the higher binding energy features in Ba 4$d$ and Y 3$d$ spectra suggests that the positive valency at these sites are getting compensated due to additional electrons near the surface area in the LSMO/YBCO sample. Such scenario is possible if there is a hole transfer across the interface to the LSMO layer as reflected in the magnetization data as also predicted in other systems \cite{STO-ALO}. Thus, the electron density in the YBCO layer will enhance which will propagate to the surface of the sample.

\section{Conclusion}

In conclusion, we have grown high quality bilayer thin films of SrTiO$_3$/YBa$_2$Cu$_3$O$_7$ and SrTiO$_3$/La$_{0.67}$Sr$_{0.33}$MnO$_3$/YBa$_2$Cu$_3$O$_7$ using a ultra high vacuum pulsed laser deposition system. The growth of both the films is found to be single crystalline, epitaxial and along (001)-direction. The in-plane(\textit{a}) and out-of-plane (\textit{c}) lattice parameters of YBCO layer matches well with the bulk orthorhombic YBCO sample. The Curie temperature of the bilayer film is significantly smaller than the Curie temperature of bulk LSMO suggesting a change in the electronic structure of the material in the bilayer. The easy axis of magnetic moment lies in the in-plane direction of LSMO layer while in the out-of-plane direction, diamagnetic behavior dominates. Unidirectional anisotropy grows with cooling and becomes large in the superconducting region.

Pristine YBCO film on STO substrate exhibit superconductivity with strong diamagnetic behavior below 86 K for the magnetic field perpendicular to the surface. The measurements for in-plane magnetic field exhibit onset of transition at a slightly higher temperature. Interestingly, LSMO/YBCO sample also show diamagnetic behavior suggesting coexistence of ferromagnetic order and superconductivity; the superconducting transition temperature remains similar despite the presence of ferromagnetic layer underneath. Ba and Y core level spectra reveals signature of different surface and bulk electronic structures. The intensity of the surface features reduce significantly in the LSMO/YBCO sample indicating an overall enhancement of electron density near the surface area. Such a scenario can appear due to the transfer of holes from YBCO to LSMO layers that also reduced magnetic transition temperature as found in magnetic measurements. A complex interplay of Meissner effect and ferromagnetic moment has been observed giving the plateau-like feature at higher applied magnetic field. These results reveal an interesting platform of coexisting ferromagnetic order and superconductivity to realize exotic science as well as advanced technology.

\section{Acknowledgements}
 Authors acknowledge the financial support from the Department of Atomic Energy (DAE), Govt. of India (Project Identification no. RTI4003, DAE OM no. 1303/2/2019/R\&D-II/DAE/2079 dated 11.02.2020). K. M. acknowledges financial support from BRNS, DAE, Govt. of India under the DAE-SRC-OI Award (grant no. 21/08/2015-BRNS/10977)


\begin{thebibliography}{99}

\bibitem{pnictides}
N. Kurita, M. Kimata, K. Kodama, A. Harada, M. Tomita, H. S. Suzuki, T. Matsumoto, Keizo Murata, S. Uji, and T. Terashima, Phys. Rev. B \textbf{83}, 214513 (2011); G. Adhikary, N. Sahadev, D. Biswas, R. Bindu,
N. Kumar, A. Thamizhavel, S. K. Dhar, and K. Maiti, J. Phys.: Condens. Matter \textbf{25}, 225701 (2013).

\bibitem{Bergeret}
F. S. Bergeret, A. F. Volkov, and K. B. Efetov, Phys. Rev. Lett. \textbf{86}, 4096 (2001).
	
\bibitem{Hoppler}
J. Hoppler, J. Stahn, Ch. Niedermayer, V. K. Malik, H. Bouyanfif, A. J. Drew, M. R\"{o}ssle, A. Buzdin,
G. Cristiani, H.-U. Habermeier, B. Keimer, and C. Bernhard, Nat. Mat. \textbf{8}, 315 (2009).
	
\bibitem{Sefrioui}
Z. Sefrioui, D. Arias, V. Pena, J. E. Villegas, M. Varela, P. Prieto, C. Leon, J. L. Martinez, and J. Santamaria, Phys. Rev. B \textbf{67}, 214511 (2003).
	
\bibitem{Djupmyr}
M. Djupmyr, S. Soltan, H.-U. Habermeier, and J. Albrecht, Phys. Rev. B \textbf{80}, 184507 (2009).
	
\bibitem{soltan2004ferromagnetic}
S. Soltan, J. Albrecht, and H.-U. Habermeier, Phys. Rev. B \textbf{70}, 144517 (2004).
	
\bibitem{Zhang}
M. J. Zhang, Y. W. Yin, T. S. Su, M. L. Teng, D. L. Zhang, X. G. Li, and L. J. Zou, Appl. Phys. Lett. \textbf{103}, 193506 (2013).
	
\bibitem{chaudhuri2023interface}
S. Chaudhuri, Y.-S. Chen, and J. G. Lin,, ACS omega \textbf{8}, 16694 (2023).

\bibitem{wisser2021growth}
Jacob J. Wisser and Yuri Suzuki, AIP Advances \textbf{11}, 015007 (2021).
	
\bibitem{chen2009flux}
C. Z. Chen, C. B. Cai, L. Peng, B. Gao, F. Fan, Z. Y. Liu, Y. M. Lu, R. Zeng, and S. X. Dou, J. Appl. Phys. \textbf{106}, 093902 (2009).
	
\bibitem{houzet2006theory}
M. Houzet and Alexandre I Buzdin, Phys. Rev. B \textbf{74}, 214507 (2006).
	
\bibitem{Samal}
D. Samal and P. S. Anil Kumar, J. Appl. Phys.\textbf{109}, 07E129 (2011).
	
\bibitem{keizer2006spin}
R. S. Keizer, S. T. B.  G\"{o}nnenwein, T. M. Klapwijk, G. Miao, G. Xiao, and A. Gupta, Nature \textbf{439}, 825 (2006).
	
\bibitem{bhatt2022correlation}
H. Bhatt, Y. Kumar, C. L. Prajapat, C. J. Kinane, A. Caruana, S. Langridge, S. Basu, and S. Singh, ACS Appl. Mater. Interfaces \textbf{14}, 8565 (2022).
	
\bibitem{hong2009vortex}
H.-Ye Wu, T. Zou, Z.-H. Cheng, and Y. Sun, Chinese Phys. Lett. \textbf{26}, 017502 (2009).
	
\bibitem{PrzyslupskiPRB}
P. Przyslupski, I. Komissarov, W. Paszkowicz, P. Dluzewski, R. Minikayev, and M. Sawicki, Phys. Rev. B \textbf{69}, 134428 (2004).
	
\bibitem{Przyslupski}
P. Przyslupski, A. Tsarou, P. Dluzewski, W. Paszkowicz, R. Minikayev, K. Dybko, M. Sawicki, B. Dabrowski, and C. Kimaball, Supercond. Sci. Technol. \textbf{19}, 538 (2006).

\bibitem{LSMO-Bindu}
J. Hemberger, A. Krimmel, T. Kurz, H.-A. Krug von Nidda, V. Yu. Ivanov, A. A. Mukhin, A. M. Balbashov, and A. Loidl, Phys. Rev. B \textbf{66}, 094410 (2002); R. Bindu, G. Adhikary, S. K. Pandey, S. Patil, K. Maiti, New J. Phys. \textbf{12}, 033026 (2010); R. Bindu, G. Adhikary, and K. Maiti, J. Phys.: Conf. Ser. \textbf{273} 012140 (2011); R. Bindu, G. Adhikary, N. Sahadev, N. P. Lalla, and K. Maiti, Phys. Rev. B \textbf{84}, 052407 (2011).

\bibitem{YBCO}
J. D. Jorgensen, B. W. Veal, A. P. Paulikas, L. J. Nowicki, G. W. Crabtree, H. Claus, and W. K. Kwok
Phys. Rev. B \textbf{41}, 1863 (1990).
	
\bibitem{Jeon}
K.-R. Jeon, C. Ciccarelli, A. J. Ferguson, H. Kurebayashi, L. F. Cohen, X. Montiel, M. Eschrig, J. W. A. Robinson, and M. G. Blamire, Nat. Mat. \textbf{17}, 499 (2018).


	
\bibitem{lorenz2001high}
M. Lorenz, H. Hochmuth, D. Matusch, M.  Kusunoki, V. L. Svetchnikov, V.  Riede, I. Stanca, G. Kastner, and D. Hesse, IEEE transactions on applied superconductivity \textbf{11}, 3209 (2001).
	
\bibitem{singh2023structural}
A. Singh, R. P. Pandeya, S. Dutta, S. C. Kandukuri, and K. Maiti, SciPost Phys. Proc. \textbf{11}, 007 (2023).
	
\bibitem{chen2000influence}
C.-H. Chen, A. Saiki, N. Wakiya, K. Shinozaki, and N. Mizutani, J. Crystal Growth \textbf{219}, 253 (2000).
	
\bibitem{liu2008microstructure}
B. Liu, R. Zhang, Z. L. Xie, H. Lu, Q. J. Liu, Z. Zhang, Y. Li, X. Q. Xiu, P. Chen, P. Han,  S. L. Gu, Y. Shi, Y. D. Zheng, and W. J. Schaff, J. Appl. Phys. \textbf{103}, 023504 (2008).

\bibitem{LSMO-bulk}
E. Dagotto, T. Hotta, and A. Moreo, Phys. Rep. \textbf{344}, 1153 (2001).

\bibitem{Balint-PNAS}
B. N\'{a}fr\'{a}di, P. Szirmai, M. Spina, A. Pisoni, X. Mettan, N. M. Nemes, L. Forr\'{o}, and E. Horv\'{a}th, PNAS \textbf{117}, 6417 (2020).

\bibitem{Tulapurkar}
Sourabh Jain, Himanshu Sharma, Amit Kumar Shukla, C. V. Tomy, V. R. Palkar, and Ashwin Tulapurkar, Physica B \textbf{448}, 103 (2014).

\bibitem{CaB6}
K. Maiti, Europhysics Letters \textbf{82}, 67006 (2008); K. Maiti, V. R. R. Medicherla, S. Patil, R. S. Singh, Phys. Rev. Letts. \textbf{99}, 266401 (2007).
	
\bibitem{kumawat2023magnetic}
S. M. Kumawat, G. D. Dwivedi, P. F. Su, W. S. Shyu, Y. H. Chien, P. W. Su, C. M. Chung, N. D. B. Fernandez, S. J. Sun, C.-H. Hsu, S. Yang, and H. Chou, J. Phys. Chem. C \textbf{127}, 6861 (2023).

\bibitem{Laverde}
M. A. Uribe-Laverde, D. K. Satapathy, I. Marozau, V. K. Malik, S. Das, K. Sen, J. Stahn, A. Ruhm, J.-H. Kim, T. Keller, A. Devishvili, B. P. Toperverg, and C. Bernhard, Phys. Rev. B \textbf{87}, 115105 (2013).

\bibitem{spectator}
R. S. Singh and K. Maiti, Phys. Rev. B \textbf{76}, 085102 (2007).

\bibitem{surface}
M. P. Seah and W. A. Dench, Surf. Interface Anal. \textbf{1}, 2 (1979); K. Maiti, P. Mahadevan, and D. D. Sarma, Phys. Rev. Lett. \textbf{80}, 2885 (1998); K. Maiti, U. Manju, S. Ray, P. Mahadevan, I. H. Inoue, C. Carbone, and D. D. Sarma, Phys. Rev. B \textbf{73}, 052508 (2006); A. Pramanik, R. P. Pandeya, K. Ali, B. Joshi, I. Sarkar, P. Moras, P. M. Sheverdyaeva, A. K. Kundu, C. Carbone, A. Thamizhavel, S. Ramakrishnan, and K. Maiti, Phys. Rev. B \textbf{101}, 035426 (2020); R. P. Pandeya, A. P. Sakhya, S. Datta, T. Saha, G. De Ninno, R. Mondal, C. Schlueter, A. Gloskovskii, P. Moras, M. Jugovac, C. Carbone, A. Thamizhavel, and K. Maiti, Phys. Rev. B \textbf{104}, 094508 (2021).

\bibitem{YBCO-hike}
K. Maiti, J. Fink, Sanne de Jong, M. Gorgoi, C. Lin, M. Raichle, V. Hinkov, M. Lambacher, A. Erb, and M. S. Golden, Phys. Rev. B  \textbf{80}, 165132 (2009).
	
\bibitem{XPS-BaO}
Osman Karslio\v{g}lu, Lena Trotochaud, Ioannis Zegkinoglou, and Hendrik Bluhm, J. Elec. Spec. Relat. Phenom. \textbf{225}, 55 (2018).

\bibitem{XPS-YBCO}
C. R. Brundle, and D. E. Fowler, Surf. Sci. Rep. \textbf{19}, 143 (1993).
	
\bibitem{XPS-Y2O3}
D. Barreca, G. A. Battiston, D. Berto, R. Gerbasi, and E. Tondello, Surf. Sci. Spectra \textbf{8}, 234 (2001).

\bibitem{STO-ALO}
N. Nakagawa, H. Y. Hwang, and D. A. Muller, Nat. Mater. \textbf{5}, 204 (2006);
S. Mukherjee, B. Pal, D. Choudhury, I. Sarkar, W. Drube, M. Gorgoi, O. Karis, H. Takagi, J. Matsuno, and D. D. Sarma, Phys. Rev. B 93, 245124 (2016).

\end{thebibliography}

\end{document}